\documentclass[9pt,twocolumn,twoside]{osajnl}

\usepackage{babel}
\newcommand{\tabincell}[2]{\begin{tabular}{@{}#1@{}}#2\end{tabular}}
\usepackage{multirow}
\usepackage{diagbox}
\usepackage{csquotes}
\journal{ol} 

\setboolean{shortarticle}{true}

\title{Frequency-resolved photon-electronic spectroscopy for excited state population detection}

\author[1]{Long Xu}
\author[2,3]{Hui Dong}
\author[2,4]{Libin Fu}

\affil[1]{Institute of Applied Physics and Computational Mathematics, Beijing 100088, China}
\affil[2]{Graduate School of China Academy of Engineering Physics, No. 10 Xibeiwang East Road, Haidian District, Beijing, 100193, China}

\affil[3]{e-mail: hdong@gscaep.ac.cn}
\affil[4]{e-mail: lbfu@gscaep.ac.cn}

\ociscodes{(020.2649) Strong field laser physics; (020.5780) Rydberg states; (300.6350) Spectroscopy, ionization; (300.6300) Spectroscopy, Fourier transforms.}

\begin{abstract}
Atomic excitation to excited states in strong laser field is the key to high-order harmonic generation below ionization threshold, yet remains unclear mainly due to the lack of proper detection methods.
We propose a frequency-resolved photon-electron spectroscopy technique to reconstruct population of excited states with the second delayed laser pulse.
The technique utilizes Fourier transformation to separate ionization from different excited states to different positions on the spectrum.
With the advantage of separation, we provide a scheme to reconstruct populations on different excited states after the first pulse.
The scheme is validated by high-precision population reconstruction of helium and hydrogen atoms.
\end{abstract}

\setboolean{displaycopyright}{true}

\begin{document}

\maketitle

The interaction of an intense laser field with atoms has led to many interesting phenomena, such as photoionization \cite{Agostini1979, Corkum1993, Becker2012, Pazourek2015, Sabbar2017, Isinger2017}, high-order harmonic generation (HHG) \cite{McPherson1987, Ferray1988, Frolov2018}, and frustrated ionization \cite{Boer1992, Nubbemeyer2008, Zimmermann2015}.
Based on the three-step model \cite{Corkum1993}, strong field approximation (SFA) can profoundly understand the physical processes that include
above-threshold ionization \cite{Wickenhauser2006,Milosevic2006}, cutoff region of HHG \cite{Lewenstein1994}, and high-energy plateau in the photoelectron spectra \cite{Becker1994, Lohr1997, Becker2002, Milosevic2009}.
However, since SFA neglects the effects of excited states, it can't describe the phenomena that involve excited states, such as Freeman resonance \cite{Freeman1987}, multiphoton Rabi oscillations \cite{Fushitani2016}, below-threshold harmonics (BTHs) \cite{Xiong2017}, and the creation of Rydberg states \cite{Zimmermann2018}.
The mechanism of BTHs has been studied with techniques such as sum frequency generation optical gating \cite{Power2010} in addition to being extensively modeled \cite{Li2015, Hassan2016}, but the role of excited state has yet to be fully examined with energy resolution.
Additionally, the observation on the effect of excited state by means of observing momentum distributions is now becoming available \cite{Krausz2009, McFarland2014}, with the development of laser technology and pump-probe spectroscopy.
Recently, in the experiment, the angular momentum component of neon atom is resolved from the photoelectron momentum distribution by the pump-probe spectroscopy \cite{Villeneuve2017}.
Moreover, the coherent dynamics of superposition of two states in hydrogen atom \cite{Zhang2017} and molecule $H_2^+$ \cite{He2018} are studied by theories.

On the experimental side, measurement methods of population on excited states are developing.
In the early days, de Boer and Muller \cite{Boer1992} used a long pulse (nanosecond probe pulse) to ionize the excited atoms and guarantee different energy peaks in energy distributions derive from different excited states.
Then, the population can be obtained by calculating the area probability of each peak.
For the metastable state, due to its long lived property, it can be detected at the microchannel plate detector (MCP) after a long flight \cite{Nubbemeyer2008, Zimmermann2018, Penent2001}.
In addition to the previous diligent effort to measure population on excited states, we propose a frequency-resolved photon-electron (FRPE) spectroscopy, which is an analogy of two-dimensional Fourier spectroscopy \cite{Aeschlimann2011, Hamm2011, Jonas2003, Oliver2014, Dong2015}
where two dimensions are all optical frequencies without the information about photoelectron momentum distribution as we have here.
In the FRPE spectroscopy, one dimension is the photoelectron momentum distribution and the other is the optical frequency.
Note that the idea of FRPE is similar to some pump probe photoelectron spectroscopy \cite{Sorensen2000, Stolow2004}, in which the delay time between pump and probe pulses are transformed to the frequency domain.
And the Fourier transform with respect to delay time has been demonstrated \cite{Frohnmeyer2000}.
With measured FRPE spectra of ejected electrons, we can reconstruct the population on excited state with a good precision.

\begin{figure}[!htb]
\centering
\includegraphics[width=7cm]{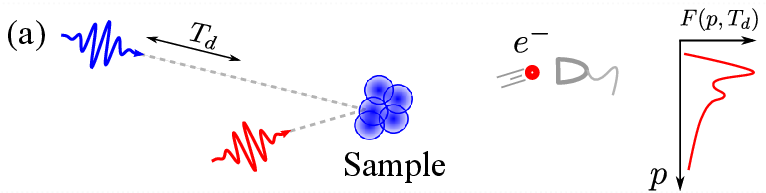}
\includegraphics[width=8cm]{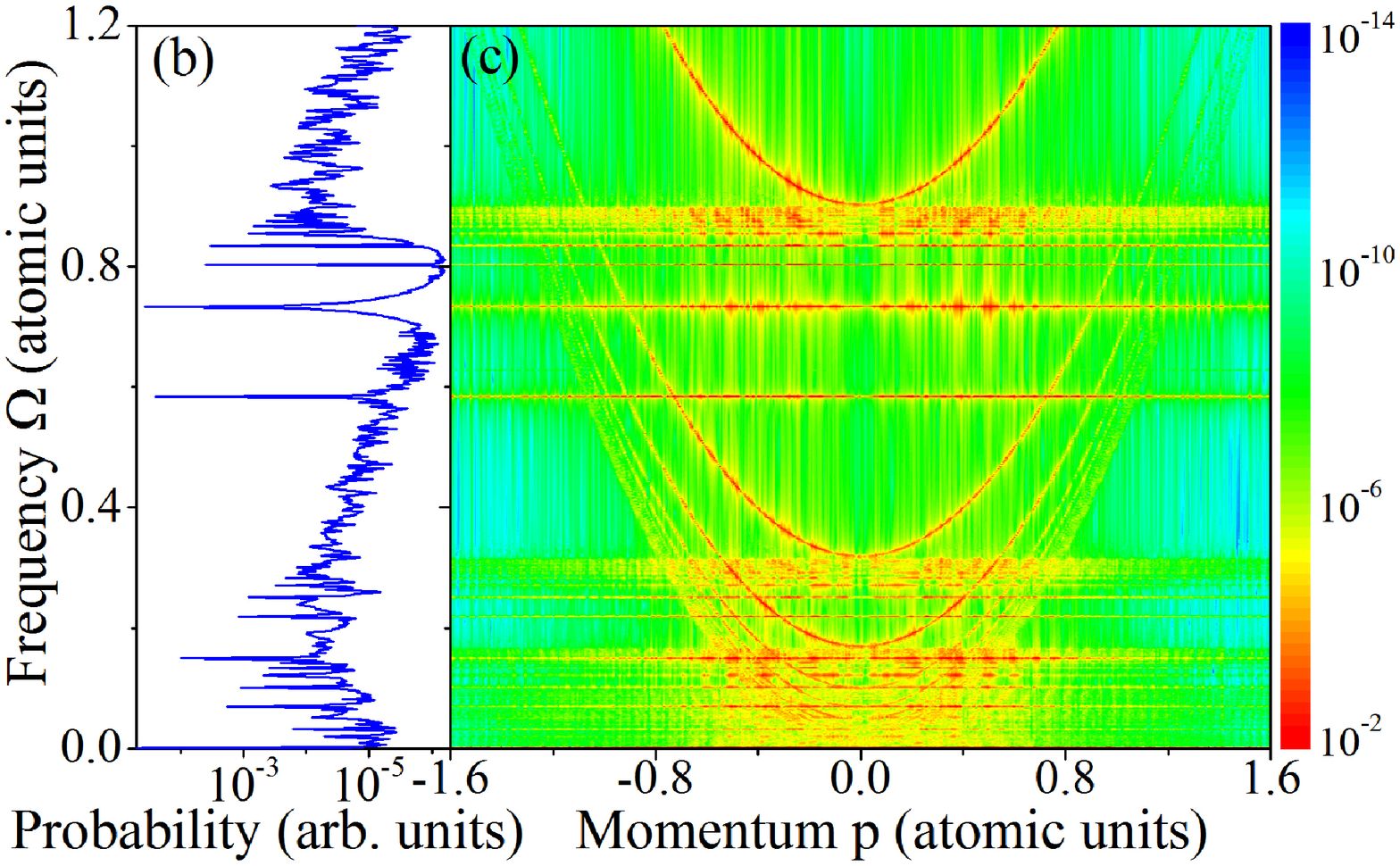}
\caption{(Color online) (a) Scheme of the simple 2-pulse frequency-resolved photon-electron spectroscopy.
The pulse 1 (in red) prepares the initial state of atoms and the pulse 2 (in blue) is used to probe the prepared state by detecting the momentum distributions of the ejected electrons.
The measurement of momentum distributions $F(p, T_d)$ is repeated with varying delay time $T_d$.
The Fourier transformation is performed for measured spectrum $F(p, T_d)$ along $T_d$ axis to obtain the final FRPE spectrum $|G(p, \Omega)|$.
A typical spectrum with Helium as the sample is shown in subfigure (c), along with the projection to the $\Omega$ axis by integration over the electron momentum $p$, namely $\int dp |G(p, \Omega)|^2$, is illustrated in (b).}
\label{fig1}
\end{figure}

As a proof-of-principle study, we design a simple experimental scheme with two subsequent laser pulses delayed by a time interval $T_{d}$,
which is defined as the time interval between the end of the first pulse and the beginning of the second pulse.
The configuration is illustrated in Fig. \ref{fig1}(a).
The first pulse is used as a pump to prepare the superposition of bound and continuum states after the atoms or molecules interact with laser field, while the second pulse, as a probing measurement, needs to be long enough to completely ionize the electrons.
A detector is set to measure the momentum distribution of the electrons.
At the overall interaction ends, we can obtain the final momentum distribution $F(p,T_{d})$ by tuning the delay time $T_{d}$.
In order to distillate the information carried by momentum distribution, we apply Fourier transform to obtain a momentum-frequency correlation spectrum $G(p,\Omega)=(2\pi)^{-1}\int F(p,T_{d})\exp(-i\Omega T_{d})dT_{d}$.
To demonstrate the advantage of the current spectroscopic method, we will numerically calculate the FRPE and show the explicit scheme to reconstruct
the population on bound states, along with the population obtained by direct calculation.

Taking helium as a sample, we show the first FRPE spectroscopy in Fig. $\ref{fig1}$ (c) and the corresponding projection to $\Omega$-axis
in Fig. $\ref{fig1}$ (b).
In the simulation, the ground state is selected as the initial state and the intensities of the two pulses are $10^{15}\mathrm{W/cm^{2}}$.
We choose 150nm laser pulse with an optical cycle and 800nm laser pulse with 64 cycles as the first and second pulses, respectively.
The specific form of laser field is given by $E(t)=-\partial_{t}A(t)$ with vector potential $A(t)=E_{0}/\omega\sin^{2}\left(\pi t/\tau\right)\cos(\omega t)$, where $E_{0}$ is the electric-field amplitude of the laser pulse with duration $\tau$  and central frequency $\omega$.
In the simulation, we obtain each slice of data $F(p,T_{d})$ by solving the time-dependent Schr\"{o}dinger equation of a single active
electron (in atomic units)
\begin{equation}
i\frac{\partial}{\partial t} \Psi(x,t)
=[-\frac{1}{2}\frac{\partial^2}{\partial x^2}-\frac{1}{\sqrt{x^2+a^2}}+x E(t,T_{d})
]\Psi(x,t),
\label{TDSE}
\end{equation}
with fixed $T_{d}$ and accumulate the full data set by scanning $T_{d}$. The second term in the right hand represents soft-Coulomb potential between electron and nucleus, where soft-core parameter \cite{Eberly1990} is chosen as $a^2=0.484$ for the singly charged helium.

To obtain the momentum distribution  $F(p,T_{d})$, we employ split-operator method \cite{feim1982} to numerically solve the Schr\"{o}dinger equation
and split the space into the inner ($|x|<x_c$) and outer ($|x|\geq x_c$) regions with $x_c$=800 a.u. by a smooth split function
\begin{equation}
F(x_c)=\left\{ \begin{array}{l}
0,\; \text{if } |x|< x_c-x_b,\\
1,\; \text{if } |x|\geq x_c,\\
\cos^4[\frac{\pi(x-x_c)}{2x_b}],\; \text{otherwise,}
\end{array}\right.
\end{equation}
where we set the width of crossover region $x_b$=50 a.u..
The wave function in the inner region $\Psi(x,t)[1-F(x_c)]$ is propagated under the full Hamiltonian numerically, while the wave function in the outer region $\Psi(x,t)F(x_c)$ is propagated under the Volkov Hamiltonian analytically \cite{Tong2006}.
In the calculation, N = 8192 grid points with a spatial step of 0.25 a.u. and the time step of 0.025 a.u. are used.
The convergence is checked by comparing with the results of the more grid points, i.e., the smaller spatial and temporal step.
Note that the selection of the spatial cutoff parameter $x_c$ is primarily determined by the amplitude of free electrons in the alternating electric field $\alpha = E_0/\omega^2$ and the convergence of xc has been confirmed numerically.
Here, we have chosen time delay from $T_d = 1000$ a.u. to $T_d = 4530$ a.u. with 2048 interval points.
The large delay time $T_d > 1000$ a.u.  ensures  enough free evolution time after the first pulse.
The Fourier transformation is performed after the simulation.
Clearly, the spectrum shows two pronounced structures: (1)  the horizontal lines at frequencies $\Omega=\mathrm{const}$, with
constants match the energy gap between two bound states, (2) the parabolic curves $\Omega=p^{2}/2+\mathrm{const}$.
Besides, the structures of the horizontal lines $\Omega=\mathrm{const}$ can also be seen from the result of frequency
distribution given by integrating the FRPE spectrum over the momentum.
We will show the physical origin of the two features in the following discussion.

 \begin{figure}[!htb]
 \centering
\includegraphics[width=7cm]{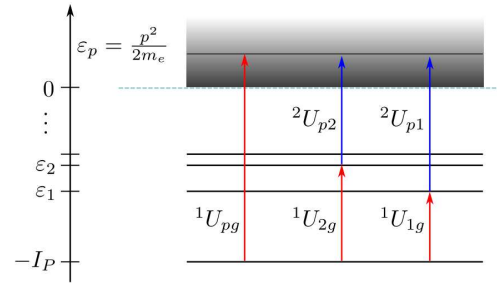}
 \caption{Ionization diagram from the perspective of energy level transitions.
g and p depict the ground and continuum states, respectively, while $\varepsilon_{j}$ labels the energy of j-th excited state.
$^{1}U$ and $^{2}U$ represent the entire system's evolutionary operator in the presence of the first and second laser pulses, respectively.}
 \label{fig2}
 \end{figure}

To understand the physical origin of the two features in the FRPE spectroscopy, we consider the evolution process from the perspective of
energy level transitions.
The atomic energy level diagram is illustrated in Fig. $\ref{fig2}$, with bound states $\{\left|e_{j}\right\rangle \} (j=0,1,2,...)$
and the continuum states $\left|p\right\rangle $.
Here $p$ is the momentum of the ionized electron.
The corresponding energy of the bound state $\left|e_{j}\right\rangle $ is denoted as $\varepsilon_{j}$.
After the first pulse, the state of the system is $\left|\Psi_{1}\right\rangle ={}^{1}U\left|\Psi_{0}\right\rangle $,
where $\left|\Psi_{0}\right\rangle =\left|g\right\rangle $ is the ground state and ${}^{1}U$ is the evolution operator of the first pulse.
In an explicit way, such state is written as $\left|\Psi_{1}\right\rangle =\sum_{j}{}^{1}U_{jg}\left|e_{j}\right\rangle +\int dp{}^{1}U_{pg}\left|p\right\rangle $, where ${}^{1}U_{jg}=\langle e_{j}|{}^{1}U|g\rangle$ and ${}^{1}U_{pg}=\langle p|{}^{1}U|g\rangle$
are the probability amplitude.
The free evolution of atom during the delay time $T_d$ induces a phase to all the states as $\left|\Psi_{1}\left(T_{d}\right)\right\rangle =\sum_{j}{}^{1}U_{jg}e^{-i\varepsilon_{j}T_{d}}\left|e_{j}\right\rangle +\int dp{}^{1}U_{pg}e^{-ip^{2}/2\ T_{d}}\left|p\right\rangle $.
After the second pulse, the wave function $|\Psi_{2}\rangle= {}^{2}U \left|\Psi_{1}\left(T_{d}\right)\right\rangle $
takes the form
\begin{equation}
|\Psi_{2}\rangle=...+\int dp(\sum_{j}{}^{2}U_{pj}{}^{1}U_{jg}e^{-i\varepsilon_{j}T_{d}}+{}^{1}U_{pg}e^{-i\frac{p^{2}}{2}T_{d}})|p\rangle,
\label{eq:Psi}
\end{equation}
where $^{2}U_{pj}=\langle p|{}^{2}U|e_{j}\rangle$ represents evolutionary operator's matrix element of the whole system when atom interacts
with the second laser pulse.
To get Eq. (\ref{eq:Psi}), we have chosen a sufficiently long $T_{d}$ to ensure that the ionized electrons obtained by the first laser pulse
will not be affected by the second pulse.
With Eq. (\ref{eq:Psi}), the amplitude of electrons in particular momentum state is
\begin{align}
\begin{split}
& F(p,T_{d})  =\left|\sum_{j}{}^{2}U_{pj}{}^{1}U_{jg}e^{-i\varepsilon_{j}T_{d}}+{}^{1}U_{pg}e^{-i\frac{p^{2}}{2} \ T_{d}}\right|^{2} \\
 & =\sum_{j}\sum_{m}|{}^{j}\Lambda_{pg}||{}^{m}\Lambda_{pg}|e^{i(\theta_{j}-\theta_{m})}
 e^{-i(\varepsilon_{j}-\varepsilon_{m})T_{d}}+|{}^{1}U_{pg}|^2\\
 & +\sum_{j}\left[{}^{j}\Lambda_{pg}{}^{1}U^*_{pg}e^{-i(\varepsilon_{j}-\frac{p^{2}}{2})T_{d}}
  +{}^{j}\Lambda^*_{pg}{}^{1}U_{pg}e^{i(\varepsilon_{j}-\frac{p^{2}}{2})T_{d}}\right],
\end{split}
\end{align}
where $^{j}\Lambda_{pg}\equiv{}^{2}U_{pj}{}^{1}U_{jg}=|{}^{j}\Lambda_{pg}|\exp(i\theta_{j})$.
The Fourier transform leads to the spectrum in the frequency domain as
\begin{align}
\begin{split}
&G(p,\Omega) =\sum_{j}\sum_{m}|{}^{j}\Lambda_{pg}||{}^{m}\Lambda_{pg}|e^{i(\theta_{j}-\theta_{m})}\delta[\Omega-(\varepsilon_{m}-\varepsilon_{j})]\\
 & +|{}^{1}U_{pg}|^2 \delta(\Omega) +\sum_{j}\left[{}^{j}\Lambda_{pg}{}^{1}U^*_{pg}\delta[\Omega+(\varepsilon_{j}-\frac{p^{2}}{2})]\right.\\
 &\left.+{}^{j}\Lambda^*_{pg}{}^{1}U_{pg}\delta[\Omega-(\varepsilon_{j}-\frac{p^{2}}{2})]\right].
\end{split}
\label{eq:Gp}
\end{align}
The first line of the above formula represents the bright lines $\Omega=\varepsilon_{j}-\varepsilon_{m}$ in the FRPE spectroscopy,
while the second and third lines show parabolic curve $\Omega=\pm(\varepsilon_{j}-p^{2}/2)$.
The current theoretical derivation shows the physical correspondence of the two main features in the Fig. \ref{fig1} (c).

The separation of different processes in the spectrum provides advantages to extract further information of the ionization process.
One application is to reconstruct the population on the bound states.
With the advantage of labeling with energy difference $\Omega_{jm}=\left|\varepsilon_{j}-\varepsilon_{m}\right|$
on the spectrum, we can get each value $|^{j}\Lambda_{pg}||{}^{m}\Lambda_{pg}|$
at $\Omega_{jm}=\left|\varepsilon_{j}-\varepsilon_{m}\right|$ ($j\neq m$) as $\left|G(p,\Omega_{jm})\right|$ .
If there are n intermediate states, we can get $C_{n}^{2}$ equations resulted from the interference between every two intermediate states.
In other words, as long as $C_{n}^{2}\geq n$, namely, $n\geq3$, we arrive at the modulus $|^{j}\Lambda_{pg}|$.
So we can get modules $|^{j}\Lambda_{pg}|$ by choosing three intermediate states and solving simultaneous equations.
Even if there are only two intermediate states ($j$ and $m$), we can solve the equations
$|^{j}\Lambda_{pg}||{}^{m}\Lambda_{pg}|=\left|G(p,\Omega_{jm})\right|$
and $|^{j}\Lambda_{pg}|^{2}+|{}^{m}\Lambda_{pg}|^{2}=\left|G(p,\Omega=0)\right|$.
For the long pulse limit to ionizing all the electrons on the bound state $\left|e_{j}\right\rangle $ via the second pulse, namely,
$\int |{}^{2}U_{pj}|^{2} dp=1$, we can get the population on the bound state $\left|e_{j}\right\rangle $ as
\begin{equation}
\begin{split}
|^{1}U_{jg}|^{2}=\int|{}^{j}\Lambda_{pg}|^{2} dp.
\end{split}
\end{equation}

\begin{table}[!htb]
\setlength{\abovecaptionskip}{0.cm}
\centering
\caption{\bf Comparison between average results obtained from the method of reconstruction with the actual value.
The reconstructed values are obtained via the reconstructed scheme.
The exact values are given by numerically solving \eqref{TDSE} for the first pulse.}
\scalebox{0.78}[0.72]{
\begin{tabular}[b]{|c|c|c|c|c|c|c|}
\hline
    \multicolumn{3}{|c|}{\diagbox[height=1.5cm,width=5.2cm]{Samples}{Population}{States}}
    & ground  & \tabincell{c}{first \\excited}  &\tabincell{c}{second \\excited}
    & \tabincell{c}{third \\excited}\\
\hline
    \multirow{6}*{He}
    & \multirow{3}*{ \tabincell{c}{150nm\\ 1cycle}}
    & Method(\%)        &   63.58   &   11.88   &   15.17   &   1.944   \\
\cline{3-7}
    &   & Exact(\%)     &   63.42   &   11.55   &   15.58   &   1.911   \\
\cline{3-7}
    & & Relative Error(\%)
                        &   0.2470  &   2.904   &   2.627   &   1.727   \\
\cline{2-7}
    & \multirow{3}*{\tabincell{c}{100nm\\ 2cycles}}
    & Method(\%)        &   45.93   &   40.56   &   7.450   &   0.3070  \\
\cline{3-7}
    &   & Exact(\%)     &   45.94   &   38.96   &   7.418   &   0.2892  \\
\cline{3-7}
    & & Relative Error(\%)
                        &   0.03193 &   4.108   &   0.4269  &   6.270   \\
\hline
\multirow{6}*{H}
    & \multirow{3}*{ \tabincell{c}{150nm\\ 1cycle}}
    & Method(\%)        &   31.82   &   54.39   &   3.578   &   2.291   \\
\cline{3-7}
    & & Exact(\%)       &   31.50   &   53.15   &   4.407   &   2.886   \\
\cline{3-7}
    & & Relative Error(\%)
                        &   1.028   &   2.329   &   18.83   &   20.63   \\
\cline{2-7}
    &\multirow{3}*{\tabincell{c}{100nm\\ 2cycles}}
    & Method(\%)        &   67.66   &   23.40   &   0.2840  &   3.231   \\
\cline{3-7}
    & & Exact(\%)       & 67.22&    22.66&  0.2840& 3.769\\
\cline{3-7}
    & &Relative Error(\%)
                        & 0.6643&   3.257&  0.00128&14.26\\
\cline{2-7}
\hline
\end{tabular}}
\label{tab:1}
\end{table}
With the reconstructed scheme, we compare the reconstructed populations on different bound states with the actual populations at the end of the first pulse from the exact numerical calculation.
The details of the comparison are shown in Table $\ref{tab:1}$ for both helium and hydrogen where the soft-core parameter in $\eqref{TDSE}$ is chosen as $a^2=2$ for the singly charged hydrogen.
Table $\ref{tab:1}$ shows the populations of four bound states for a total of four samples including two atoms and two laser pulses.
Here we only choose 4 lowest levels as an example, the distribution of the remaining populations is mainly in the continuum states and partly in higher lying bound states.
Owing to the property of Discrete Fourier Transform, the frequency gap after Fourier Transform is $\Delta\Omega =2\pi/(N \triangle T_d)=2\pi/(4530-1000)=0.00178$ a.u., where N and $\triangle T_d$ represent the number of sampling points and sampling interval, respectively.
Additionally, the results of reconstruction are the average of $C_4^3 =4$ situations because we can solve the populations for every three states.
In the simulation, we do not consider any decay of the bound states, noticing that the lifetime ($10^{-9}$s) of excited states in
atoms and molecules is far longer than the time scale of the current scheme ($10^{-13}$s).
From the table, we know that all the results of more populations are accurate, where most of the relative errors are below $5\%$.
Moreover, the reconstruction of hydrogen is not as accurate as helium.
The only reason is the ionization potential of H ($I_P=0.5 a.u.$) is smaller than He ($I_P=0.9 a.u.$),
leading to less effective points $I_P/\Delta\Omega$ in the range of the ionization potential.

\begin{figure}[!htb]
\centering
\includegraphics[width=\linewidth]{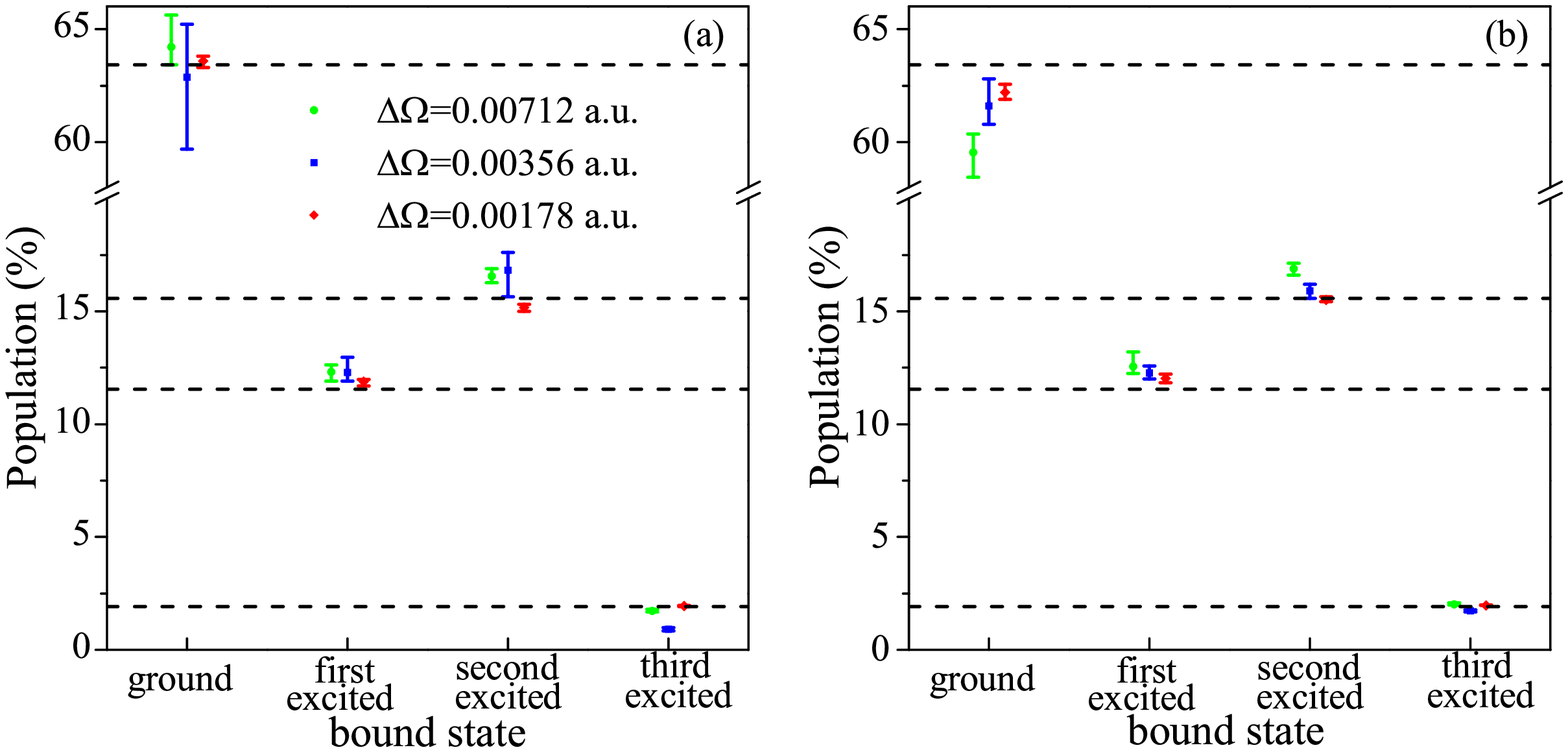}
\caption{The reconstructed populations of the first sample for three different frequency gap $\Delta\Omega$,
0.00712 (green), 0.00356 (blue), 0.00178 (red) a.u..
The points represent the average values, while the error bar represents the maximum and minimum values.
The dashed lines depict the exact values given by numerically solving \eqref{TDSE}.
The modulus $\left|G(p,\Omega_{jm})\right|$ is chosen as the maximum value in (a) and the mean value in (b).}
\label{fig3}
\end{figure}

In view of the deviation of the result, we further study its accuracy.
Taking the first sample as an example, namely, helium atom exposed to a 150nm laser pulse with 1 optical cycle, we plot the
reconstructed populations in Fig. $\ref{fig3}$ (a) where
the modulus $\left|G(p,\Omega_{jm})\right|$ is chosen as the maximum value in the
neighbor of $\Omega_{jm}=\left|\varepsilon_{j}-\varepsilon_{m}\right|$.
Obviously, the populations of $\Delta\Omega=0.00178$ a.u. are more concentrated
around the actual value than $\Delta\Omega=0.00712$ a.u., whereas the results of $\Delta\Omega=0.00356$ a.u. are
spread out, as shown in Fig. $\ref{fig3}$ (a).
This is entirely due to the fact that the function $\delta[\Omega-(\varepsilon_{m}-\varepsilon_{j})]$ is discrete rather than continuous.
Additionally, taking the area of the $\left|G(p,\Omega)\right|^2$ instead of a point, we reselect $\left|G(p,\Omega_{jm})\right|$ as the 0.5th power of the sum of the square of a left point and a right point of $\Omega_{jm}$, that is, $\left|G(p,\Omega_{jm})\right| = \sqrt{\left|G(p,\Omega_{jm\mathrm{left}})\right|^2 + \left|G(p,\Omega_{jm\mathrm{right}})\right|^2 }$.
The corresponding reconstructed populations are plotted in Fig. $\ref{fig3}$ (b).
Compared with the results of Fig. $\ref{fig3}$ (a), the populations are more concentrated around the average values.
More importantly, the accuracy is getting better with the decreasing $\Delta\Omega$.
Consequently, it's indeed an effective way to improve accuracy by increasing the effective delay
time length to reduce the frequency gap $\Delta\Omega$.

In summary, we designed the frequency-resolved photon-electron spectroscopy to detect the population on the excited states for atoms under impact of the strong laser pulse.
With the numerical simulation, we showed the typical FRPE spectrum  with two main features for helium, and explained their physical origins.
With the spectrum, we presented the detailed scheme to reconstruct population, and validated the scheme with Helium as the example.
The current scheme shows a good agreement between reconstructed value and the exact value. However, to experimentally demonstrate the Fourier transform, it's important to establish the phase stability between two pulses. Currently, the stable phase between different color pulses in the experiment \cite{Oliver2014, Dong2015, Konar2018} have been achieved for weak fields.

{\bf Funding.}
National Natural Science Foundation of China (Grant No. 11725417, 11575027, 11875049), NSAF (Grant No. U1730449), and Science Challenge Project (Grant No. 2018005); The Recruitment Program of Global Youth Experts of China.

\newpage

\renewcommand\refname{Full References}

\end{document}